\documentclass[sigconf,authorversion,nonacm]{acmart}

\usepackage{subcaption}
\usepackage{savesym}
\usepackage{listings}
\usepackage[show]{chato-notes}
\usepackage{amsmath}

\usepackage{amssymb}%
\usepackage{pifont}%

\newcommand{\cmark}{\ding{51}}%
\newcommand{\xmark}{\ding{55}}%

\newcommand{\pageenlarge}[1]{}

\usepackage{marginnote}

\newcommand{\ca}[1]{\textcolor{black}{#1}} %
\newcommand{\maria}[1]{\textcolor{black}{#1}}
\newcommand{\cm}[1]{\textcolor{black}{#1}} %
\newcommand{\craig}[1]{\textcolor{black}{#1}} %

\begin{document}

\title{On Coherence-based Predictors for Dense Query Performance Prediction}

\author{Maria Vlachou}
\affiliation{%
  \institution{University of Glasgow}
  \city{Glasgow}
  \country{UK}
}
\email{m.vlachou.1@research.gla.ac.uk}

\author{Craig Macdonald}
\affiliation{%
  \institution{University of Glasgow}
  \city{Glasgow}
  \country{UK}
}
\email{craig.macdonald@glasgow.ac.uk}

\begin{abstract}
\looseness -1 Query Performance Prediction (QPP) estimates the effectiveness of a search engine's results in response to a query without relevance judgments. Traditionally, {\em post-retrieval} predictors have focused upon either the distribution of the retrieval scores, or the coherence of the top-ranked documents using traditional bag-of-words index representations. More recently, BERT-based models using dense embedded document representations have been used to create new predictors, but mostly applied to predict the performance of rankings created by BM25. Instead, we aim to predict the effectiveness of rankings created by single-representation dense retrieval models (ANCE \& TCT-ColBERT). Therefore, we propose a number of variants of existing unsupervised coherence-based predictors that employ neural embedding representations. In our experiments on the TREC Deep Learning Track datasets, we demonstrate improved accuracy upon dense retrieval (up to 92\% compared to sparse variants for TCT-ColBERT and 188\% for ANCE). Going deeper, we select the most representative and best performing predictors to study the importance of differences among predictors and query types on query performance. Using existing distribution-based evaluation QPP measures and a particular type of linear mixed models, we find that query types further significantly influence query performance (and are up to 35\% responsible for the unstable performance of QPP predictors), and that this sensitivity is unique to dense retrieval models. Our approach introduces a new setting for obtaining richer information on query differences in dense QPP that can explain potential unstable performance of existing predictors and outlines the unique characteristics of different query types on dense retrieval models.
\end{abstract}
\maketitle

\section{Introduction}
Retrieval effectiveness in search engines can vary across different queries~\cite{harman2004nrrc,voorhees2003overview}. \ca{\cm{Being able to accurately predict} the likely effectiveness of a search engine for \cm{a given query} may facilitate interventions, such as asking the user to reformulate the query.} \cm{To this end, the} task of {\em Query Performance Prediction (QPP)} aims to predict the effectiveness of a search result in response to a query without having access to relevance judgments~\cite{carmel2010estimating}.  In the last two decades, a number of {\em query performance predictors} \ca{have been} proposed, which can be grouped in two main categories: {\em Pre-retrieval} predictors estimate query performance using only linguistic or statistical information contained in the queries or the corpus~\cite{he2004inferring,hauff2008survey,zhao2008effective,scholer2009case,mothe2005linguistic}. On the other hand, {\em post-retrieval} predictors use the relevance scores \ca{or contents} of the top returned documents, by measuring, for example, the focus of the result list compared to the corpus~\cite{cronen2002predicting,zhou2007query}, or the distribution of the scores of the top-ranked documents~\cite{cummins2011improved,perez2010standard,roitman2017robust,shtok2009predicting,tao2014query}. Predictors based on NQC~\cite{shtok2012predicting} (standard deviation of relevance scores) have been found to be quite promising. A further group of predictors examine the pairwise relations among the retrieved documents~\cite{arabzadeh2021query, diaz2007performance}. Thus far, these predictors have been applied using   traditional bag-of-words \cm{representations}. While examining the coherence between returned documents is useful, as we show, these representations are not suitable for predicting the query performance of more advanced retrieval methods.

\looseness -1 More recently, pre-trained language models (PLMs) have introduced neural network architectures that encode the embeddings of queries and documents~\cite{devlin2018bert,khattab2020colbert,lin2020distilling,xiong2020approximate}, and, as a consequence, increase retrieval effectiveness by retrieving the documents whose embeddings are spatially similar to the query embeddings. Often, a BERT-based model is trained for use as a reranker of the result retrieved by (e.g.) BM25~\cite{robertson1994some} - such {\em cross-encoders} include  BERT\_CLS~\cite{nogueira2019passage} and monoT5~\cite{nogueira2020document}. \cm{On the other hand}, {\em dense retrieval approaches}~\cite{karpukhin2020dense,xiong2020approximate} are increasingly popular, whereby an embedding-based representation of the documents are indexed, and those with the \cm{similar embeddings} to the query are identified through nearest-neighbour search (e.g.\ ANCE~\cite{xiong2020approximate}, TCT-ColBERT~\cite{lin2020distilling}). Compared to reranking setups, dense retrieval is attractive as recall is not limited by the initial BM25 retrieval approach, and improvements in the PLM can improve the retrieval effectiveness. \maria{Therefore, dense retrieval models inspire us to develop predictors that are effective for predicting their rankings.}

\pageenlarge{1} In parallel, neural architectures have also seen some adoption as methods for predicting query difficulty. These post-retrieval methods are {\em supervised}, and use refined neural architectures in order to produce a final performance estimate~\cite{arabzadeh2021bert,datta2022pointwise,hashemi2019performance,zamani2018neural}. For instance, BERT-QPP~\cite{arabzadeh2021bert} fine-tunes the BERT language model~\cite{devlin2018bert} embeddings to the QPP task by \cm{estimating the relevance of the top-ranked document retrieved for each query.} Still, BERT-based supervised QPP approaches were initially evaluated of in terms of their correlation with the effectiveness of a BM25 ranking. However, it was recently shown that the performance of BERT-QPP is significantly lower or is outperformed by unsupervised predictors when using more advanced retrieval methods and the TREC Deep Learning datatsets~\cite{faggioli2023query}. Instead, we believe that less expensive unsupervised predictors that take advantage of the nature of advanced retrieval methods could be more indicative of query performance.

\looseness -1 \maria{Therefore, in this work, we aim to accurately predict the effectiveness of recent dense retrieval methods. In this regard, we extend current predictors and propose ways to improve their performance under the new models. We reason that the confidence of well-trained and effective ranking methods should be indicative of the likely effectiveness of the system for a given query.  Therefore, we consider patterns among the document embeddings of the retrieved documents and revisit existing unsupervised coherence-based predictors ~\cite{arabzadeh2021query,diaz2007performance}. More importantly, we update the type of representations used by these predictors from traditional sparse to dense representations that match those of the corresponding retrieval method. In this way, we manage to create applicable predictors for dense retrieval. In addition, we study how the performance of each query varies among all applicable predictors, and how this is further influenced by differences among query types.}

\maria{Indeed, while it is widely known that query effectiveness is different across queries~\cite{harman2004nrrc,voorhees2003overview}, it is relatively unknown how QPP varies across different query types. For example, Carmel at al.~\cite{carmel2006makes} studied the components related to single topic difficulty, while Faggioli and Marchesin~\cite{faggioli2021makes} explored which queries in the medical domain are suited for lexical or semantic retrieval models and label queries into semantically easy or hard. Still, these has been considerably less work on how QPP is related to taxonomies of what types (groups) of questions are generally asked. Recently, a categorisation method has been proposed~\cite{bolotova2022non}, which assigns questions in existing datasets to a pre-defined set of categories. In this taxonomy, the questions belonging to certain categories were found to be more difficult to answer compared to others. In this regard, the extent to which QPP varies according to existing query taxonomies has not received much attention. Indeed, knowing which types of queries are more difficult to answer would inform us about how to develop predictors that are specifically designed for them. In particular, it would be useful to examine the performance of different predictors and see how each is or is not affected by categorisation.}

In short, our contributions are the following: (i) We propose a number of embedding variants of existing coherence predictors and our own extension {\em pairRatio}, an unsupervised predictor which extends the intuition of spatial autocorrelation~\cite{diaz2007performance} and uses pairwise relations of embedding vectors. In this way, we create predictors designed for dense retrieval; (ii) We extensively study existing predictors to two state-of-the-art single-representation dense retrieval models, namely ANCE~\cite{xiong2020approximate} and TCT-ColBERT~\cite{lin2020distilling}, as well as BM25, and show that changing the representations increases performance significantly not just for dense but also sparse retrieval; (iii) By also comparing with supervised predictors, we show that applying a BERT-based model for QPP might be a misdirection of resources - applying a BERT model would be preferable for improving the effectiveness of the ranking; (iv) By applying statistical analyses on query performance, we propose a new way of measuring and splitting the total variation in QPP performance into within-query and between-query importance. At the same time, we detect a unique sensitivity of dense retrieval methods, which are affected by query type (up to 35\% increase in query performance variations due to query categorisation) and exhibit larger differences between predictors, a pattern which is not apparent in sparse retrieval.
The structure of the rest of this paper is as follows: We present related work in Section~\ref{sec:rw}, and present our new extended predictors in Section~\ref{sec:meth}. Then, we follow with traditional correlation analysis of QPP predictors in Sections~\ref{sec:exp_setup} and~\ref{sec:results}, continue with an extended linear mixed model analysis to test for query type in Section~\ref{sec:stat}, and conclude with some final remarks in Section~\ref{sec:conc}.

\pageenlarge{1}\section{Related Work}\label{sec:rw}
\maria{The focus of this paper is on post-retrieval QPP, as post-retrieval QPPs are in general more accurate than pre-retrieval QPPs~\cite{hauff2008survey}. Indeed, there are two main reason why we eliminate pre-retrieval predictors from our focus. First, existing recent unsupervised neural pre-retrieval predictors propose, for example,  geometric semantic similarities of query terms, which indicate query specificity and are based on pre-trained neural embeddings~\cite{arabzadeh2020neural,roy2019estimating}. Since these predictors examine queries at the token-level, they are not applicable to single-representation dense retrieval models. Second, information based on queries can, in general, provide quite limited information with respect to the effectiveness of the ranking. Therefore, we focus on predictors that examine the top-returned list of documents.}

\maria{In terms of post-retrieval QPP, earlier post-retrieval predictors examined} the focus on the result list induced by language models (probability distributions of all single terms)~\cite{cronen2002predicting}. For example, {\em Clarity}~\cite{cronen2002predicting} measures the divergence of the language model of top-ranked documents from the one of the corpus. \maria{The intuition is that the corpus is considered as an irrelevant set, and the higher the divergence from it, the better the performance.} {\em Utility Estimation Framework (UEF)}~\cite{shtok2010using} uses pseudo-effective reference lists induced by language models. \maria{These are language models based on term probabilities, and are therefore, different from neural PLMs mentioned in Section 1} and estimates their relevance using score-based predictors \maria{such as NQC (see below in Section~\ref{ssec:score_qpp})}. Both of these rely upon term probabilities, and are, therefore, not feasible for extending our predictions to dense retrieval. {\em Query Feedback (QF)}~\cite{zhou2007query} refers to the overlap of the returned documents with those obtained after applying pseudo-relevance feedback - yet, pseudo relevance feedback approaches for dense retrieval are still in their infancy~\cite{wang2021pseudo,yu2021improving}, so we do not consider Query Feedback further. 

\craig{In the remainder of this section, we discuss the main types of query performance predictors that could be applied to dense retrieval, specifically score-based unsupervised predictors (Section~\ref{ssec:score_qpp}) and document representation-based predictors (Section~\ref{ssec:emb_qpp}).}

\pageenlarge{1}\subsection{Score-based QPP}\label{ssec:score_qpp}
\looseness -1 Score-based predictors encode certain assumptions about how the scores should be distributed for high or low-performing queries. For instance, a simple predictor might be  the {\em Maximum Score} among the retrieved documents~\cite{roitman2017enhanced} - the higher the maximum score, the more confident the retrieval system is that it has found a document that matches well the query. Still, the maximum score is not often used without normalisation, as the scores produced by term weighting model models – such as TF.IDF and BM25 – vary depending on the query length. The most commonly applied score-based predictor is {\em Normalised Query Commitment (NQC)}~\cite{shtok2009predicting}, which is based on the standard deviation of the retrieval scores, which is negatively correlated with the amount of query drift (the non-related information in the result list)~\cite{mitra1998improving}. Several variations of standard deviation have been proposed in the literature that aim to further enhance its accuracy~\cite{cummins2011improved,perez2010standard}. Other predictors extend the intuition of NQC by incorporating the scores magnitude~\cite{tao2014query}, or by estimating a more robust version of variance resulting from bootstrap samples that represent a population of scores~\cite{roitman2017robust}. Indeed, {\em Robust Standard Deviation estimator (RSD)}~\cite{roitman2017robust} is an example of extending the results to multiple contexts represented by bootstrap samples. Score-based predictors are easily applicable to dense retrieval, since scores are computed by each retrieval method. %

\subsection{Document Representation-based QPP}\label{ssec:emb_qpp}
Document representations provide richer information about documents, since they capture their semantic underlying information~\cite{devlin2018bert,lin2020distilling}. Therefore, \maria{since predictors based on document representations capture semantic relations either between queries, documents, or their interaction}, these predictors are directly applicable to dense retrieval models. 

\subsubsection{Unsupervised Coherence Predictors}
In general, effective unsupervised predictors that consider document representations are preferable, since they require less computation \maria{than supervised predictors}. One \maria{example of an unsupervised predictor} that examines the lexical representations of documents is {\em spatial autocorrelation}~\cite{diaz2007performance}, which considers the spatial proximity of lexical document representations, by using their pairwise TF.IDF-based similarities to produce a new set of scores ``diffused in space''. The final predictor is obtained by correlating the original scores with the diffused scores. \craig{Indeed, a low correlation between scores of topically-close documents is assumed to imply a poor retrieval performance.}

Another family of recent coherence-based predictors \cm{creates a graph of the most similar documents among the top-ranked documents~\cite{arabzadeh2021query}, based on their TF-IDF representations}.
Specifically, metrics such as Weighted Average Neighbour Degree (WAND) and Weighted Density (WD) were found to enhance the performance of score-based predictors after linear interpolation. While these predictors use pairwise similarities, they have thus far been applied to sparse document representations and therefore, not previously applied to dense retrieval.

\pageenlarge{1}\subsubsection{Supervised \& Neural Predictors}
\maria{In general, supervised models for QPP can be attractive due to the varying sources of indicators for inferring query performance~\cite{roitman2017enhanced}. However, at the same time, they bring more computational complexity compared to unsupervised predictors.} For example, Neural-QPP~\cite{zamani2018neural} is a multi-component supervised predictor as the output of existing unsupervised QPP predictors with weak supervision \cm{- we can think of this as a neural supervised aggregation predictor}. %

\cm{More recently}, BERT-QPP~\cite{arabzadeh2021bert} was proposed, which \cm{fine-tunes a BERT model for} the QPP task by adding cross-encoder or bi-encoder \cm{layers that estimate an effectiveness measure (e.g. NDCG) based on the contents of the top returned document in response to the query. While BERT-QPP can also be applied to the ranking returned by a dense retrieval approach, it uses a different model instance to that used by the dense retrieval approach itself.} \maria{Out of the two BERT-QPP variants, the bi-encoder version is closer to the intuition of single-representation dense retrieval.} Finally, Datta et al. proposed qppBERT-PL~\cite{datta2022pointwise}, which adds an LSTM network on top of the BERT representation to model both document contents and the progression of estimated relevance in the ranking. \cm{Compared to BERT-QPP, this approach has promise as it considers more information than just the top-ranked document.}

\maria{To summarise, existing predictors have either focused on sparse document representations or retrieval scores on the unsupervised side, or have introduced neural pre-trained architectures to create more complex supervised predictors.} \craig{However, no work has addressed unsupervised predictors using dense embedded representations, as are readily available in dense retrieval configuration}. Indeed, if we want to predict the ranking of a more effective dense retrieval method, using a simple predictor that takes into account document relations or simply the scores returned by a dense method could be sufficient for QPP, without the need to apply an additional cross-encoder-based supervised prediction model. In the next section, \craig{we describe in detail} existing predictors that can be applied to dense retrieval.

\section{Coherence Predictors for Dense Retrieval}\label{sec:meth}
\maria{In this section, we first describe some existing sparse coherence-based predictors in Section~\ref{ssec:meth_sp}, and then show how these can be adapted to be better suited for dense retrieval settings in Section~\ref{ssec:meth_den}.}

\subsection{Sparse Coherence-based Methods}\label{ssec:meth_sp}

\maria{Here, we start with some existing sparse coherence-based query performance prediction methods, namely {\em spatial autocorrelation}~\cite{diaz2007performance}, and two network metrics, WAND and WD~\cite{arabzadeh2021query}. }

\subsubsection{Spatial Autocorrelation (AC)}
First, consider $d$ to be a document's TF.IDF vector. Then, the inner product of two documents at ranks $i$ and $j$ is given by $sim(d_i,d_j)$. We can obtain a pairwise similarity matrix as follows:
\begin{equation}\label{eq_2}
W = 
\begin{bmatrix}
sim(d_{11}) & sim(d_{12}) & ... & sim(d_{1k})\\
... & ... & ... & ...\\
sim(d_{k1}) & sim(d_{k2}) & ... & sim(d_{kk})
\end{bmatrix}
\end{equation}
where $k$ is the cutoff number of the top-k documents. For brevity of notation, let $sim(d_{ij}) = sim(d_i,d_j)$. Projecting (multiplying) each element of the matrix $W_{ij}$ on the vector of the original retrieved scores, $Score(\vec{d})$, \cm{we can} obtain a new list of {\em diffused} scores as:
\begin{equation}\label{eq_3}
Score(\tilde{d}) = W *  Score(d)   
\end{equation}
Thereafter, an estimate of the spatial autocorrelation (AC)~\cite{diaz2007performance} can be obtained by using the Pearson correlation between the two vectors: 
\begin{equation}\label{eq_4}
AC = corr(Score(\tilde{d}), Score(d))  
\end{equation}
which quantifies the relation between the initial and diffused scores.  \ca{Indeed, as mentioned above, a low correlation between the original retrieval scores (i.e. $Score(d)$) and those weighted by their topical similarity (the diffused scores, $Score(\tilde{d})$) was found to imply poor retrieval performance~\cite{diaz2007performance}.}

\subsubsection{Network Metrics}
\maria{As mentioned above, the matrix $W$ represents all pairwise similarities between the top-retrieved documents. This matrix is equivalent to a fully connected network, where each node $\mathbf{V_G}$ corresponds to the $d$  TF.IDF vector, and each edge $\mathbf{E_G}$ corresponds to each entry $sim(d_{ij})$~\cite{arabzadeh2021query}, or more formally $\mathbf{G}(q, D_q^{(k)}) = {\{\mathbf{V_G, E_G},W\}}$.} In this regard, to avoid all edges being considered equal without attention to the edge weight, the network is further pruned via thresholding~\cite{christophides2015entity}, where the similarities higher than the mean similarity value are selected as neighbours.

Consequently, we have the following definitions, which correspond to some recently proposed network metrics~\cite{arabzadeh2021query} for QPP:
\begin{equation}\label{eq_7}
Average Neighbour Degree (AND) = \frac{1}{k} \sum_{i=1}^{k} (\frac{1}{n_i}{\sum_{j\in N_{d_i}} n_j} )
\end{equation} 
where $N_{d_i}$ is the neighbourhood of document $i$ and $n_i$ is the number of its neighbours. \maria{Actually, Equation~\eqref{eq_7} corresponds to the Weighted AND measure or {\em WAND}~\cite{arabzadeh2021query}, as AND is applied on the pruned graph that only contains edges between the most similar documents.}

\maria{Another way to think about coherence is to count the observed edges or similarities over the set of all possible edges. This results in the Density measure, as follows:}
\begin{equation}\label{eq_8}
Density (D) = \frac{2|\mathbf{E_G}|}{|\mathbf{V_G}|(|\mathbf{V_G}|-1)}
\end{equation}
In short, a higher neighbourhood degree and a higher density of a graph network indicates a more coherent set of top-retrieved results. The general intuition behind these measures is that the presence of coherence, as reflected by highly similar documents in a top-retrieved set indicates the ability of the retrieval method to distinguish relevant from non-relevant documents, and therefore, return the relevant ones at the top of the list.

\pageenlarge{1}\subsection{Dense Coherence-based Methods}\label{ssec:meth_den}

\begin{figure}[tb]
\begin{subfigure}{0.20\textwidth}
\centering
\includegraphics[width=\textwidth]{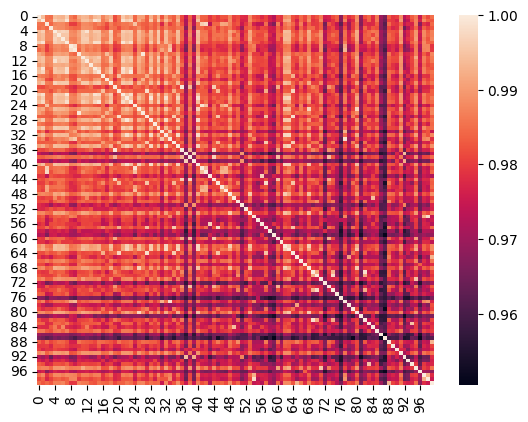}
\caption{Effective Query}\label{fig:mat1}
\end{subfigure}
\begin{subfigure}{0.20\textwidth}
\centering
\includegraphics[width=\textwidth]{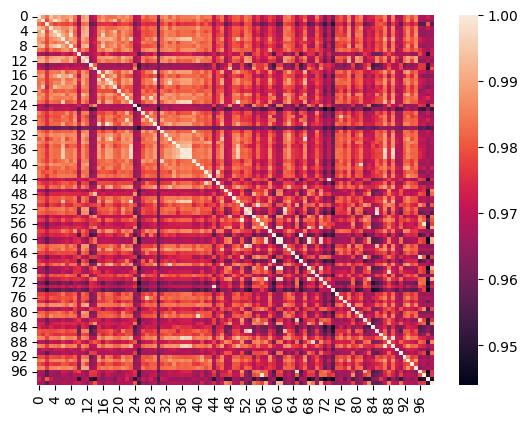}
\caption{Ineffective Query}\label{fig:mat2}
\end{subfigure}\vspace{-1em}
\caption{Heatmap of pairwise similarity matrix of the top-100 TCT-ColBERT document embeddings for returned for the best (query id 104861 with NDCG@10=1) and worst performing queries (query id 489204 with NDCG@10=0.189) from the TREC DL 19 queryset.}\label{fig:mat}\vspace{-\baselineskip}
\end{figure}

We now derive the embedding representation variants of the above predictors in order to make them suitable for the prediction of neural dense retrievers. We first create the variants for embedding-based AC and network metrics, and then introduce a new variant that extends AC by considering rank groupings.

\subsubsection{AC-embs}
Let $\phi_d$ and $\theta_q$ respectively represent the dense embedded representation of a document and a query. Firstly, we adapt autocorrelation, such that instead of TF.IDF vectors we consider the embedded document representations. Let the inner product of two documents at ranks $i$ and $j$  (with embeddings $\phi_{i}$ and $\phi_{j}$) be written $sim(\phi_{dij})$, then we can define the pairwise similarities of the top $k$ ranked documents as:
\begin{equation}\label{eq_5}
W^{\phi} = 
\begin{bmatrix}
sim(\phi_{d11}) & sim(\phi_{d12}) & ... & sim(\phi_{d1k})\\
... & ... & ... & ...\\
sim(\phi_{dk1}) & sim(\phi_{dk2}) & ... & sim(\phi_{dkk})
\end{bmatrix}
\end{equation}
We can then apply autocorrelation (denoted as AC above) as per Equations~\eqref{eq_3} \& \eqref{eq_4}. We denote this as {\em AC-embs}. 

\subsubsection{Network-embs}
Similarly, and as we showed that the similarity matrix is equivalent to a fully connected network set of edges, we can apply WAND and WD as per Equations~\eqref{eq_7} \& \eqref{eq_8}, denoted as {\em WAND-embs} and {\em WD-embs}, respectively.

\subsubsection{pairRatio}
We now introduce an extension of AC-embs inspired by visually exploring embedding relations. Specifically, in Figure~\ref{fig:mat}, we visualise the pairwise similarity matrix ($W_{\phi}$) obtained using TCT-ColBERT~\cite{lin2020distilling} embeddings for the top-100 passages for the one high and one low performing query in the TREC Deep Learning Track 2019 queryset. For the best performing query, there is higher pairwise similarity among documents of top ranks (top left corner, indicated by a group of lighter shading), and lower correlation for lower ranks (darker shading). On the other hand, for the worst query, elements of darker shading appear at high ranks, indicating that the top-ranked documents may not be as coherent). \maria{In addition, there is less dark shading in low ranks compared to the best query.} These observations inspire us to explore the trend of average top vs.\ bottom rank pairwise similarities of top-ranked embeddings.

Specifically, let $W^{\phi}_{\tau_1..\tau_2}$ denote the (diagonal) subset of $W^{\phi}$ between ranks $\tau_1$ and $\tau_2$. Then, for a given rank threshold $\tau$, we can measure the ratio between the mean pairwise similarity above and below rank $\tau$, \cm{i.e. $W^{\phi}_{0..\tau}$ and $W^{\phi}_{\tau..k}$} as follows:
\begin{equation}
pairRatio(W^{\phi}) = (\overline{W^{\phi}_{1..\tau}}) \cdot (\overline{W^{\phi}_{\tau..k}})^{-1} \label{eqn:pr}
\end{equation}
where $\overline{W^{\phi}}$ denotes the mean of the given matrix. 
In practice, we separately adjust the end of the upper matrix and the start of the lower matrix as separate hyperparameters. We called this predictor {\em pairRatio}. \maria{Unlike WAND and WD, we consider the magnitude of this contrast as indicative of query performance. We believe that, since this relates to the retrieval method itself, it should be indicative of query performance especially for advanced retrieval methods.}

Still, the similarity matrix $W^{\phi}$ can only provide information about the relative similarity of documents. Introducing some information about the document scores would increase performance prediction accuracy, \maria{since it relates to the absolute ranking of each document}. Let $A$ be an adjusted matrix, where each entry, for a document pair $i$ and $j$ is multiplied by the final similarity of the query to each of the documents:
$$
A_{ij} = W_{ij} \cdot (\phi_i \cdot \theta_q)\cdot  (\phi_j \cdot \theta_q)
$$
$A$ better encodes similarity of the query among the pairwise document similarities. pairRatio (Equation~\eqref{eqn:pr}) can then be applied upon $A$, which we denote as adjusted pairRatio, or {\em A-pairRatio}.

\maria{In short, we are interested in the effectiveness of these predictors based on dense document representations and how they perform in relation to their sparse versions. We test their performance compared to score-based and supervised predictors in Section~\ref{sec:results}. Later, we analyse the prediction results using Linear Models to check what further characteristics influence query performance in Section~\ref{sec:stat}. We next describe the experimental setup for our experiments.}

\section{Experimental Setup}\label{sec:exp_setup}

Our experiments address the following research questions:

{\bf RQ1} How do unsupervised coherence-based predictors compare to unsupervised score-based predictors in dense and sparse retrieval?

{\bf RQ2} How do unsupervised predictors perform compared to supervised predictors in dense and sparse retrieval?

\noindent To address these research questions, our setup is as follows:

{\bf Datasets:} We conduct experiments using the MSMARCO passage ranking corpus, and applying the TREC Deep Learning track 2019 and 2020 query sets, containing respectively 43 and 54 queries with relevance judgements. In particular, each query in these querysets contains many judgements obtained by pooling various distinct retrieval systems.

{\bf QPP Predictors}: As unsupervised \craig{score-based} predictors, we apply Max score (MAX)~\cite{roitman2017enhanced}, and NQC~\cite{shtok2009predicting}. As a representative variant of NQC, we choose {\em RSD}. This bootstrap-based predictor is the most recent NQC variant and was shown to outperform other score-based predictors. Specifically, we use the {\em RSD(uni)} version \craig{which samples documents uniformly}.  For each cutoff, we sample from 0.60 to 0.80 of the initial result list size.  We use spatial autocorrelation (AC)~\cite{diaz2007performance}, WAND and WD~\cite{arabzadeh2021query}, and the interpolation of WAND and WD with NQC. We use interpolation following the findings of the original paper~\cite{arabzadeh2021query}, which suggest that network metrics further increase the performance of NQC. We also report our embedding variants (AC-embs, WAND-embs, WD-embs, PairRatio, A-PairRatio). For each unsupervised predictor, we tune the hyperparameters of each dataset on the other. Specifically, to tune the cutoff value for the top-$k$ documents all unsupervised predictors including ours, we use a grid of values [5,10,20,50,100,200,500,1000]. For PairRatio and A-PairRatio, we also vary the other upper and lower rank threshold hyperparameter $\tau$.%

For supervised predictors, we report the bi-encoder and cross-encoder variants of BERT-QPP~\cite{arabzadeh2021bert}. To achieve this, we retrained the BERT-QPP cross-encoder and bi-encoder models specifically for each of the dense retrieval models. These supervised predictors exhibit their highest correlations mainly for MRR, which means that they train models that estimate the relevance of the top document of a ranking. In this regard, we check whether an alternative supervised predictor (which we call {\em top-1(monoT5)}) that uses only the top-retrieved document to a monoT5 model~\cite{nogueira2020document} -- \craig{i.e.\ trained for relevance estimation and ranking rather than performance prediction} --  can perform well in dense retrieval. Note that we deliberately use the term {\em QPP Predictors} instead of {\em baselines}, since our purpose is not to demonstrate the superiority of a single predictor, but rather how a {\em group of predictors} behaves under different contexts and retrieval models.

{\bf Retrieval Systems}: We deploy three retrieval approaches: BM25 sparse retrieval (applying Porter's English stemmer and removing standard stopwords) as implemented by Terrier~\cite{ounis06terrier-osir}, and two single-representation dense retrieval approaches, namely ANCE~\cite{xiong2020approximate}, and TCT-ColBERT~\cite{lin2020distilling} with PyTerrier~\cite{macdonald2021pyterrier} integrations.\footnote{https://github.com/terrierteam/pyterrier\_dr} 

{\bf Measures}: Following the TREC 2019 Deep Learning Track Over\-view~\cite{craswell2020overview}, we measure system effectiveness in terms of NDCG@10 and MAP@100. We further add MRR@10, following some recent work~\cite{arabzadeh2021bert,hashemi2019performance}. To quantify the accuracy of the QPP techniques, we adopt Kendall's $\tau$ correlation measure, as typically reported in QPP literature~\cite{cronen2002predicting,diaz2007performance,hauff2008survey,shtok2009predicting,shtok2010using,shtok2012predicting,zamani2018neural}.\footnote{We omit Pearson's correlation, as it assume a linear relationship between variables.}

\begin{table*}[tb]
\caption{Kendall's $\tau$ correlations of unsupervised and supervised predictors for TREC DL 2019. The highest correlation by an unsupervised predictor in each column is emphasised in bold and (*) indicates significance at $\alpha = 0.05$.}\label{tab:dl19}
\begin{tabular}{c|ccccccccc}
\hline
                  & \multicolumn{3}{c}{BM25}                                                                                             & \multicolumn{3}{c}{ANCE}                                                                                             & \multicolumn{3}{c}{TCT}                                                                         \\ \hline
                  & {\color[HTML]{333333} MAP@100} & {\color[HTML]{333333} NDCG@10} & \multicolumn{1}{c|}{{\color[HTML]{333333} MRR@10}} & {\color[HTML]{333333} MAP@100} & {\color[HTML]{333333} NDCG@10} & \multicolumn{1}{c|}{{\color[HTML]{333333} MRR@10}} & {\color[HTML]{333333} MAP@100} & {\color[HTML]{333333} NDCG@10} & {\color[HTML]{333333} MRR@10} \\ \hline
Effectiveness     & 0.232                          & 0.479                          & \multicolumn{1}{c|}{0.639}                         & 0.332                          & 0.643                          & \multicolumn{1}{c|}{0.806}                         & 0.402                          & 0.720                          & 0.898                         \\ \hline
& \multicolumn{9}{c}{Score-based} \\ \hline
Max               & 0.171                          & 0.157                          & \multicolumn{1}{c|}{0.087}                         & 0.428*                         & 0.316*                         & \multicolumn{1}{c|}{0.241*}                        & 0.297*                         & 0.250*                         & 0.015                         \\
NQC               & 0.322*                         & 0.281*                         & \multicolumn{1}{c|}{0.075}                         & {\bf 0.499*}                         & 0.463*                         & \multicolumn{1}{c|}{0.216}                         & {\bf 0.335*}                         & 0.243*                         & 0.171                         \\
RSD(uni)          & 0.328*                         & 0.288*                         & \multicolumn{1}{c|}{0.077}                         & 0.495*                         & {\bf 0.467*}                         & \multicolumn{1}{c|}{0.264*}                        & {\bf 0.335*}                         & 0.228*                         & {\bf 0.227}                        \\ \hline
& \multicolumn{9}{c}{Sparse Coherence-based} \\ \hline
AC                & 0.156                         & 0.073                         & \multicolumn{1}{c|}{0.071}                        & 0.111                        & 0.081                        & \multicolumn{1}{c|}{0.061}                        & 0.080                        & -0.198                        & -0.051                         \\
WAND              & 0.209*                         & 0.126                          & \multicolumn{1}{c|}{0.111}                         & 0.187                          & 0.113                          & \multicolumn{1}{c|}{0.025}                         & 0.189                          & 0.095                          & -0.006                        \\
WD                & 0.158                          & 0.101                          & \multicolumn{1}{c|}{0.087}                         & 0.158                          & -0.004                         & \multicolumn{1}{c|}{-0.009}                        & 0.184                          & 0.121                          & 0.015                         \\
WAND(NQC)         & 0.258*                         & 0.148                          & \multicolumn{1}{c|}{0.124}                         & 0.178                          & 0.113                          & \multicolumn{1}{c|}{0.025}                         & 0.189                          & 0.095                          & -0.01                         \\
WD(NQC)           & 0.200*                         & 0.186                          & \multicolumn{1}{c|}{0.035}                         & 0.158                          & -0.008                         & \multicolumn{1}{c|}{-0.012}                        & 0.18                           & 0.135                          & 0.006                         \\ \hline
& \multicolumn{9}{c}{Dense Coherence-based} \\ \hline
WAND-embs         & -0.096                         & -0.232                         & \multicolumn{1}{c|}{-0.019}                        & 0.138                          & -0.157                         & \multicolumn{1}{c|}{-0.029}                        & -0.036                         & 0.139                          & 0.041                         \\
WD-embs           & 0.224*                         & -0.170                         & \multicolumn{1}{c|}{0.014}                         & 0.089                          & -0.219                         & \multicolumn{1}{c|}{-0.241*}                       & -0.147                         & -0.033                         & 0.045                         \\
AC-embs           & 0.373*                         & 0.144                          & \multicolumn{1}{c|}{0.098}                         & 0.437*                         & 0.285*                         & \multicolumn{1}{c|}{0.261*}                        & 0.056                          & 0.018                          & -0.129                        \\
pairRatio(ours)   & 0.171                          & 0.270*                         & \multicolumn{1}{c|}{0.194}                         & 0.295*                         & 0.334*                         & \multicolumn{1}{c|}{0.087}                         & 0.200                          & 0.248*                         & -0.060                        \\
A-pairRatio(ours) & {\bf 0.446*}                         & {\bf 0.352*}                         & \multicolumn{1}{c|}{0.142}                         & 0.382*                         & 0.403*                         & \multicolumn{1}{c|}{0.216}                         & 0.280*                         & {\bf 0.259*}                         & 0.171                         \\ \hline
& \multicolumn{9}{c}{Supervised} \\ \hline
BERT-QPP (bi)     & 0.229*                         & 0.305*                         & \multicolumn{1}{c|}{0.260*}                        & 0.162                          & 0.144                          & \multicolumn{1}{c|}{0.067}                         & 0.111                          & 0.048                          & 0.083                         \\
BERT-QPP(cross)   & 0.264*                         & 0.254*                         & \multicolumn{1}{c|}{0.174*}                        & 0.198                          & 0.117                          & \multicolumn{1}{c|}{0.038}                         & 0.211*                         & 0.088                          & 0.041                         \\ 
top-1(mono-T5)   & 0.180                         & 0.294*                         & \multicolumn{1}{c|}{\bf 0.359*}                        & 0.224*                          & 0.294*                          & \multicolumn{1}{c|}{\bf 0.470*}                         & 0.058                         & 0.038                          & 0.086                         \\ \hline
\end{tabular}
\end{table*}

\begin{table*}[tb]
\caption{Results on TREC DL 2020. Notation as per Table~\ref{tab:dl19}.}
\begin{tabular}{c|ccccccccc}
\hline
                  & \multicolumn{3}{c}{BM25}                                                                                             & \multicolumn{3}{c}{ANCE}                                                                                             & \multicolumn{3}{c}{TCT}                                                                         \\ \hline
                  & {\color[HTML]{333333} MAP@100} & {\color[HTML]{333333} NDCG@10} & \multicolumn{1}{c|}{{\color[HTML]{333333} MRR@10}} & {\color[HTML]{333333} MAP@100} & {\color[HTML]{333333} NDCG@10} & \multicolumn{1}{c|}{{\color[HTML]{333333} MRR@10}} & {\color[HTML]{333333} MAP@100} & {\color[HTML]{333333} NDCG@10} & {\color[HTML]{333333} MRR@10} \\ \hline
Effectiveness     & 0.275                          & 0.493                          & \multicolumn{1}{c|}{0.614}                         & 0.363                            & 0.607                          & \multicolumn{1}{c|}{0.803}                         & 0.454                          & 0.686                          & 0.831                         \\ \hline
& \multicolumn{9}{c}{Score-based} \\ \hline
Max               & 0.215*                         & 0.214*                         & \multicolumn{1}{c|}{0.184}                         & 0.213*                         & 0.285*                         & \multicolumn{1}{c|}{0.337*}                        & 0.342*                         & 0.243*                         & 0.062                         \\
NQC               & 0.526*                         & 0.438*                         & \multicolumn{1}{c|}{0.281*}                        & {\bf 0.443*}                         & 0.082                          & \multicolumn{1}{c|}{0.172*}                        & {\bf 0.454*}                         & 0.246*                         & 0.133                         \\
RSD(uni)          & 0.568*                         & 0.431*                         & \multicolumn{1}{c|}{0.288*}                        & 0.403*                         & 0.275*                         & \multicolumn{1}{c|}{0.155}                         & 0.335*                         & 0.341*                         & 0.208*                        \\ \hline
& \multicolumn{9}{c}{Sparse Coherence-based} \\ \hline
AC                & -0.199*                        & 0.017                        & \multicolumn{1}{c|}{-0.097}                        & -0.115                         & -0.022                         & \multicolumn{1}{c|}{-0.014}                       & 0.018                          & -0.118                          & 0.030                        \\
WAND              & 0.189*                         & -0.031                         & \multicolumn{1}{c|}{-0.026}                        & 0.130                          & 0.009                          & \multicolumn{1}{c|}{-0.065}                        & 0.208*                         & 0.220*                         & 0.023                         \\
WD                & 0.183*                         & 0.006                          & \multicolumn{1}{c|}{-0.036}                        & 0.158                          & 0.044                          & \multicolumn{1}{c|}{0.01}                          & 0.225*                         & 0.216*                         & 0.018                         \\
WAND(NQC)         & 0.220*                         & 0.101                          & \multicolumn{1}{c|}{-0.024}                        & 0.130                          & 0.005                          & \multicolumn{1}{c|}{-0.067}                        & 0.202*                         & 0.213*                         & 0.188                         \\
WD(NQC)           & 0.253*                         & 0.160                           & \multicolumn{1}{c|}{0.036}                         & 0.148                          & 0.023                          & \multicolumn{1}{c|}{-0.010}                        & 0.223*                         & 0.192*                         & 0.004                         \\ \hline
& \multicolumn{9}{c}{Dense Coherence-based} \\ \hline
WAND-embs         & 0.038                          & 0.137                          & \multicolumn{1}{c|}{0.042}                         & 0.291*                         & {\bf 0.300*}                         & \multicolumn{1}{c|}{0.077}                         & -0.05                          & 0.107                          & -0.066                        \\
WD-embs           & 0.099                          & 0.158                          & \multicolumn{1}{c|}{0.028}                         & 0.213*                         & 0.289*                         & \multicolumn{1}{c|}{\bf 0.394*}                        & 0.127                          & 0.127                          & -0.161                        \\
AC-embs           & {\bf 0.607*}                         & {\bf 0.443*}                         & \multicolumn{1}{c|}{0.339*}                        & 0.324*                         & 0.219*                         & \multicolumn{1}{c|}{0.149}                        & 0.121                          & 0.137                          & -0.002                        \\
pairRatio(ours)   & 0.271*                         & 0.203*                          & \multicolumn{1}{c|}{0.130}                         & 0.178                          & 0.186                          & \multicolumn{1}{c|}{-0.132}                        & 0.364*                         & 0.318*                         & {\bf -0.280*}                       \\
A-pairRatio(ours) & 0.482*                         & 0.316*                         & \multicolumn{1}{c|}{0.189}                         & 0.348*                         & 0.270*                         & \multicolumn{1}{c|}{0.115}                         & 0.429*                         & {\bf 0.363*}                         & { -0.244*}                        \\ \hline
& \multicolumn{9}{c}{Supervised} \\ \hline
BERT-QPP (bi)     & 0.322*                         & 0.315*                         & \multicolumn{1}{c|}{0.351*}                        & 0.274*                         & 0.047                          & \multicolumn{1}{c|}{0.058}                         & 0.353*                         & 0.195*                         & 0.083                         \\
BERT-QPP(cross)   & 0.375*                         & 0.345*                         & \multicolumn{1}{c|}{0.403*}                        & 0.180                          & 0.043                          & \multicolumn{1}{c|}{0.012}                         & 0.261*                         & 0.173                          & 0.041                         \\ 
top-1(mono-T5)   & 0.371*                         & 0.400*                         & \multicolumn{1}{c|}{\bf 0.534*}                        & 0.259*                          & 0.237*                          & \multicolumn{1}{c|}{0.365*}                         & 0.279*                         & 0.240*                          & 0.159                         \\ \hline
\end{tabular}
\end{table*}

\section{Correlation Results}\label{sec:results}
Tables 1 and 2 show the accuracy of all our examined predictors on the TREC DL 2019 and 2020 query sets, respectively. Within each table: groups of columns denote the various retrieval approaches; the uppermost row reports the mean effectiveness of each ranking approach for each evaluation measure; the next group of rows contains the Kendall's $\tau$ correlation of the score-based predictors, the next one the unsupervised lexical coherence-based predictors; then we report the results for the embedding-based predictors; and finally for the supervised predictors~\cite{arabzadeh2021bert}.

\subsection{RQ1: Score-based vs Coherence-based Predictors}
As expected, for BM25, distribution-based score predictors (NQC and RSD(uni) show high accuracy for MAP@100 and NDCG@10, while their accuracy is lower for MRR@10, especially for DL 19. However, unlike older datasets, sparse coherence predictors are very low for TREC DL datasets. As for dense coherence predictors, surprisingly, AC-embs variant is the best performing predictor for AP@100, and for NDCG@10 on 2020. As for our pairRatio variants, they are less effective than other unsupervised predictors, such as NQC and AC-embs (except for MRR@10), as well as supervised predictors on MRRR@10. 

\cm{Next we consider the} two dense retrieval settings, \cm{i.e. ANCE \& TCT-ColBERT}. For TCT-ColBERT,  we observe that our pairRatio predictors outperform not only supervised predictors, but also NQC (the best performing unsupervised predictor) for NDCG@10 and MRR@10 for both datasets, are only behind RS(uni) for MRR@10 in the DL 2019 dataset, and are competitive for AP@100. Another observation is that A-pairRatio has increased the accuracy compared to pairRatio, particularly for the TCT-ColBERT model, which indicates the need for including document-query relations.
For ANCE, WAND-embs and WD-embs are better than score-based predictors for NDCG@10 and MRR@10 for the 2020 datatset, while they are only slightly behind them in the 2019 dataset.
Overall, for MAP@100, NQC or RSD (uni) consistently outperform coherence-based predictors, while for NDCG@10 and MRR@10, the picture is more unstable; however, in most cases, coherence-based predictors win for dense retrieval. \craig{Further, as might be expected,} changing the type of representations from sparse to dense increases the performance of coherence-based predictors across the dense retrieval settings, as the updated representations match those of the retrieval methods. To answer RQ1, for dense retrieval, score-based predictors perform well for MAP@100, while coherence-based predictors show increased accuracy for NDCG@10 and MRR@10. For sparse retrieval, dense coherence predictors are in general better than score-based predictors.

\pageenlarge{1} \subsection{RQ2: Unsupervised vs. Supervised Predictors}
Next, we compare the performance of unsupervised with supervised QPP predictors for each retrieval method. For BM25, we are able to reproduce the results of the bi-encoder and cross-encoder variants of BERT-QPP, \maria{as reflected by the higher values in MRR and the competitive correlation on the other two metrics. For  BM25, we used the authors' checkpoints, while we re-trained the method for dense retrieval.} However, their values are still lower than NQC, \maria{(a simple score-based unsupervised predictor),} and RSD(uni) (NDCG@10 on the TREC 2019 queryset), our pairRatio (MRR@10 on the 2019 queryset), AC-embs (AP@100 on 2019, AP@100 on 2020, NDCG@10 on 2020), and top-1 monoT5 (MRR@10 on both datasets). Most importantly, for the two dense retrieval methods, they are not as effective as unsupervised predictors, such as Max and NQC. For TCT-ColBERT, supervised predictors are less effective than our pairRatio variants for NDCG@10 and MRR@10, and NQC and RSD(uni) for all metrics. The \cm{strongest observed} correlations of BERT-QPP variants in dense retrieval are for AP@100. However, \cm{they have a cost to deploy} (applying a BERT model on the top-ranked result), which could be instead used to re-rank the top results. In addition, the simpler "supervised" variant, {\em top-1(mono-T5),} which uses the monoT5 score of the top-ranked document is a more accurate predictor than BERT-QPP across all retrieval methods, particularly for MRR@10, \maria{which is the metric that BERT-QPP is most competitive}. This surprising result shows that BERT-QPP is itself just a relevance estimator for the top-ranked document that has been trained to predict MRR@10; using any effective relevance estimator can do as good a job, if not better. 

To answer RQ2, we find that the existing BERT-QPP supervised predictors are less accurate than unsupervised predictors (existing and ours) for dense retrieval.

\section{Linear Mixed Effects Models for QPP}\pageenlarge{1}\label{sec:stat}
\subsection{Modeling Query Differences in QPP}\label{ssec:mod}
From the above results, we observe the following: First, the performance of dense representation coherence-based predictors is quite unstable across metrics, as seen, for example, from its lower performance on MAP@100 compared to the other measures. However, coherence-based predictors are demonstrated to be particularly useful in certain dense retrieval settings (for TCT-ColBERT: pairRatio and A-pairRatio, for ANCE: WAND-embs and WD-embs). Second, we see that none of the examined predictors can outperform all other predictors consistently across all settings. 

Therefore, we ask: (1) What makes the QPP predictors unstable? It could be that the correlation between a predictor's estimate and the observed effectiveness of the query is further influenced by the type of query (i.e., Experience, which was found difficult to answer~\cite{bolotova2022non}). In this case, we need to determine the importance of query type and how it interacts with each QPP predictor; (2) Can we make inferences about the presence of significant differences between different QPP predictors? While predictors differ in terms of Kendall's $\tau$, these differences are not, in most cases, extremely large. We can test this using a statistical model that compares the effectiveness of the different predictors. 

In this regard, Faggioli et al.~\cite{faggioli2021enhanced} proposed a QPP evaluation approach based on the scaled Absolute Rank Error (sARE). Specifically, sARE for each query is defined as: $sARE_{q_{i}} = \frac{|r^p_i - r^e_i|}{|Q|}$
where $r^p_i$ and $r^e_i$ are the ranks assigned to query $i$ by the QPP predictor and the evaluation metric, respectively. In this way, measuring QPP accuracy moves from a point estimate prediction approach, as provided by the correlation, to a distribution-based approach, where one sARE value is obtained per query. 

This further allows to model sARE as a function of other factors in Analysis of Variance (ANOVA) models~\cite{faggioli2021enhanced,faggioli2023query}. However, we need to make one important distinction. On one hand, each query receives a separate sARE value for every QPP predictor (within-query factor). On the other hand, each query belongs to only one of the levels of query type as proposed in ~\cite{bolotova2022non}. We, therefore, treat query type as a between-query factor. Next, we show how the within- and between-query levels can be modeled separately and provide their relative contribution.

In our analysis, we only include predictors that are most representative for dense retrieval: NQC, RSD(uni), Max, dense coherence-based predictors, and supervised predictors. To properly match the type of predictor representations with the ones of the retrieval models, for BM25, we use the sparse version of coherence predictors, as they were originally used for BM25 rankings\footnote{As a sanity check, we also conducted a separate analysis using the dense coherence predictors to obtain sARE from BM25, and the results were similar.}. For each selected predictor, we use the sARE values obtained from the predictor values and their corresponding evaluation from Section 5. We merge the two TREC DL query sets, and each query is assigned to one of the following categories or, as we call them, {\em query types}: Evidence-based, Factoid, Experience, Instruction, Reason, and Not a Question~\cite{bolotova2022non}.

\begin{table}[ht]
\caption{Interpretation of terms included in the linear mixed effects full model.}
\resizebox{85mm}{!}{
\begin{tabular}{cl}
\hline
\multicolumn{1}{c|}{Parameter}     & \multicolumn{1}{c}{Interpretation}                                                                           \\ \hline
\multicolumn{2}{c}{Fixed effects}                                                                                                                 \\ \hline
\multicolumn{1}{c|}{$\gamma_{00}$} & \begin{tabular}[c]{@{}l@{}}average true sARE for the reference QPP predictor for \\ the reference (without the effect of) query type\end{tabular} \\
\multicolumn{1}{c|}{$\gamma_{01}$} & \begin{tabular}[c]{@{}l@{}}average difference in sARE between different query\\ types for the reference QPP predictor\end{tabular}    \\
\multicolumn{1}{c|}{$\gamma_{10}$} & \begin{tabular}[c]{@{}l@{}}average true rate of change in sARE per unit change \\ in QPP predictor for the reference (without the effect of)\\ query type\end{tabular}   \\
\multicolumn{1}{c|}{$\gamma_{11}$} & \begin{tabular}[c]{@{}l@{}}average difference in sARE between different query \\ types per unit change in QPP predictor\end{tabular}   \\ \hline
\multicolumn{2}{c}{Random effects}                                                                                                                \\ \hline
\multicolumn{1}{c|}{$\zeta_{0i}$, $\zeta_{1i}$}  &\begin{tabular}[c]{@{}l@{}}allow individual true query trajectories to be scattered\\ around the average query true change trajectory\end{tabular}                                                                                                              \\
\multicolumn{1}{c|}{$\epsilon_{ij}$} &\begin{tabular}[c]{@{}l@{}}allows individual query data to be scattered around\\ individual query true change trajectory\end{tabular}                                                                                                              \\ \hline
\multicolumn{2}{c}{Variance Components}                                                                                                           \\ \hline
\multicolumn{1}{c|}{$\sigma^2_\epsilon$}  & \begin{tabular}[c]{@{}l@{}}level 1 (residual) variance, variability around each\\ query's true change trajectory \end{tabular}                                                                                           \\
\multicolumn{1}{c|}{$\sigma^2_0$, $\sigma^1_1$} &  \begin{tabular}[c]{@{}l@{}}level 2 variance in reference predictor and rate of\\ change per predictor measurement, how much between-\\query variability is left after accounting for query type \end{tabular}                                                                                                             \\
\multicolumn{1}{c|}{$\sigma_{01}$} &\begin{tabular}[c]{@{}l@{}}residual covariance between true sARE for the reference\\ (initial) predictor and rate of change, controlling for\\ query type, across all queries \end{tabular}                                                                                                              \\ \hline
\end{tabular}}
\end{table}

\begin{table*}[tb]
\caption{Resulting LME models for each retrieval method and all metrics.}
\resizebox{\textwidth}{!}{
\begin{tabular}{cc}
\hline
\multicolumn{2}{c}{BM25}                                         \\ \hline
\multicolumn{1}{l|}{$sARE_{MAP}$} & $sARE_{ij} = [0.29 - 0.009(QPP Predictor_{ij})] + [\zeta_{0i} + \zeta_{1i}(QPP Predictor_{ij}) + \epsilon_{ij}]$                     \\
\multicolumn{1}{l|}{$sARE_{NDCG}$} & $sARE_{ij} = 0.26 + \zeta_{0i} +  \epsilon_{ij}$                     \\
\multicolumn{1}{l|}{$sARE_{MRR}$}  & $sARE_{ij} = 0.30 + \zeta_{0i} +  \epsilon_{ij}$                     \\ \hline
\multicolumn{2}{c}{ANCE}                                         \\ \hline
\multicolumn{1}{l|}{$sARE_{MAP}$} & $sARE_{ij} = [0.28 -0.008(QPP Predictor_{ij}) + 0.25(NotAQ_i) + 0.05(NotAQ_i)(QPP Predictor_{ij})] +  [\zeta_{0i} + \zeta_{1i}(QPP Predictor_{ij}) + \epsilon_{ij}]$                     \\
\multicolumn{1}{l|}{$sARE_{NDCG}$} & $sARE_{ij} = 0.25 + \zeta_{0i} + \epsilon_{ij}$                     \\
\multicolumn{1}{l|}{$sARE_{MRR}$}     & $sARE_{ij} = [0.35 - 0.008(QPP Predictor_{ij})] + [\zeta_{0i} + \zeta_{1i}(QPP Predictor_{ij}) + \epsilon_{ij}]$                      \\ \hline
\multicolumn{2}{c}{TCT-ColBERT}                                  \\ \hline
\multicolumn{1}{l|}{$sARE_{MAP}$} & $sARE_{ij} = [0.32 - 0.01(QPP Predictor_{ij}) + 0.05(Experience_i)(QPP Predictor_{ij})] +  [\zeta_{0i} + \zeta_{1i}(QPP Predictor_{ij}) + \epsilon_{ij}]$    \\
\multicolumn{1}{l|}{$sARE_{MAP}$} & $sARE_{ij} = [0.32 - 0.01(QPP Predictor_{ij}) + 0.02(Reason_i)(QPP Predictor_{ij})] +  [\zeta_{0i} + \zeta_{1i}(QPP Predictor_{ij}) + \epsilon_{ij}]$                     \\
\multicolumn{1}{l|}{$sARE_{NDCG}$} & $sARE_{ij} = [0.32 - 0.008(QPP Predictor_{ij})] + [\zeta_{0i} + \zeta_{1i}(QPP Predictor_{ij}) + \epsilon_{ij}]$                     \\
\multicolumn{1}{l|}{$sARE_{MRR}$}  & $sARE_{ij} = 0.32 + \zeta_{0i} +  \epsilon_{ij}$                      \\ \hline
\end{tabular}}
\end{table*}

\begin{table*}[ht]
\caption{Proportion of explained variance per component and included fixed effects in each LME for all three retrieval methods. \cmark indicates the presence of a fixed effect in LMEs, while \xmark shows the absence of either an important contribution of a factor (top) or a fixed effect (bottom).}
\begin{tabular}{c|ccc|ccc|ccc}
\hline
   & \multicolumn{3}{c|}{BM25}                             & \multicolumn{3}{c|}{ANCE}                               & \multicolumn{3}{c}{TCT-ColBERT}                            \\ \hline
   & {\ $sARE_{MAP}$} & $sARE_{NDCG}$ & $sARE_{MRR}$ & { $sARE_{MAP}$} & $sARE_{NDCG}$ & $sARE_{MRR}$   & {$sARE_{MAP}$} & $sARE_{NDCG}$   & $sARE_{MRR}$    \\ \hline
$Pseudo-R^2_\epsilon$ & 13.4\%    &  \xmark    &     \xmark    &7.5\%   & \xmark & 16.5\%     & 12.4\% & 14.6\%  &\xmark\\
$Pseudo-R^2_0$ &  \xmark    & \xmark     &\xmark     & 17.2\%  &\xmark      & \xmark      & 2.2\%                      & 9.9\%  & \xmark \\
$Pseudo-R^2_1$ & \xmark     &  \xmark   &  \xmark   & 35.6\%   & \xmark     & \xmark      & 22.8\%                    & 8.1\%  & \xmark  \\ \hline
$\gamma_{00}$  & \cmark          & \cmark     &\cmark  & \cmark      &  \cmark    & \cmark      & \cmark                                         &\cmark        & \cmark       \\
$\gamma_{01}$  & \xmark          & \xmark     & \xmark    & \cmark   & \xmark     &  \xmark     & \cmark                              & \xmark       &\xmark        \\
$\gamma_{10}$  &  \cmark         &  \xmark    & \xmark    & \cmark   &  \xmark    & \cmark      & \xmark                             &  \cmark      &  \xmark      \\
$\gamma_{11}$  & \xmark          &  \xmark    & \xmark    & \cmark   &  \xmark    &\xmark       &  \cmark                              & \xmark       &  \xmark      \\ \hline
\end{tabular}
\end{table*}

\subsection{Linear Mixed Model Definitions}
In this section, \craig{we describe the} {\em Linear Mixed Effects (LME)} models~\cite{curran1997relation,field2012discovering,maxwell2017designing,singer2003applied} applied on QPP. LMEs are part of the Generalised Linear Models (GLM)~\cite{madsen2010introduction,nelder1972generalized} and allow us to determine the proportion of within-and between query variations. While GLMs have recently been proposed in information retrieval for measuring differences between systems ~\cite{faggioli2022detecting}, they have received much less attention in QPP, and have only recently been used in the form of ANOVA~\cite{faggioli2021enhanced,faggioli2023query}. Unlike Faggioli et al.~\cite{faggioli2023query}, we apply our GLMs on each retrieval method separately, and split the total variation in $sARE$ into within-query and between-query levels. This allows us to differentiate between the effects of QPP predictors and those of query types on sARE. In the following, using a model selection strategy, we show how much is gained in explained variance when adding a factor in each of the two levels. Level 1 specifies the within-query variations over different QPP predictors (how each query changes over different QPP measurements). Level 2 specifies the differences in changes associated with query type (between-query differences). In other words, Level 2 further specifies the intercept and slope terms of Level 1 by including, for each, an effect of the between-query factor (in our case: query type). Table 3 shows the interpretation of each of the full model parameters. Next, we describe our examined LMEs.

First, the full model for predicting sARE, denoted as $LME_{full}$, can be defined as follows: 

\noindent {\bf Level 1}
\begin{equation}\label{eq_full_lme}
sARE_{ij} = \pi_{0i} + \pi_{1i}(QPP Predictor) + \epsilon_{ij}
\end{equation}
with $\epsilon_{ij}\sim~N(0, \sigma^2_\epsilon)$

\noindent where $sARE_{ij}$ is the sARE value of query $i$ at QPP predictor measurement $j$, $\pi_{0i}$ is the intercept (initial status) of query $i$'s change trajectory (in our case, it is the reference QPP predictor, specifically the first QPP measurement), $\pi_{1i}$ is the slope (rate of change) in sARE (per unit of measurement), and $\epsilon_{ij}$ are the deviations of a query's equation on each measurement.

\noindent {\bf Level 2}
\begin{equation}\label{eq_full_lme_2}
   \begin{cases}
\pi_{0i} = \gamma_{00} + \gamma_{01}(Query Type) + \zeta_{0i} \\
\pi_{1i} = \gamma_{10} + \gamma_{11}(Query Type) + \zeta_{1i} 
   \end{cases}
\end{equation}

with
$\begin{matrix} \zeta_{0i} \\ \zeta_{1i} \end{matrix}
 \sim~MVN
  \begin{bmatrix}
      \begin{bmatrix}
       0 \\
       0
      \end{bmatrix},
      \begin{bmatrix}
       \sigma^2_0  \sigma_{01} \\
       \sigma_{01}  \sigma^1_1
      \end{bmatrix}
   \end{bmatrix}$

\noindent Specifically, $\gamma_{00}$ and $\gamma_{10}$ are the average true sARE without the effect of query type (for the query type that is set as a reference category) in the initial status and rate of change, respectively. Similarly, $\gamma_{01}$ and $\gamma_{11}$ show the effect of the between-query factor, namely query type on sARE, for the initial status and rate of change. For convenience, we will use $LME_{full}$ in its equivalent composite form as:
\begin{equation}\label{eq_full_comp}
   \begin{split}
sARE_{ij} = [\gamma_{00} + \gamma_{10}(QPP Predictor_{ij}) + \gamma_{01}(Query Type_i)\\ + \gamma_{11}(Query Type_i)(QPP Predictor_{ij})] \\+  [\zeta_{0i} + \zeta_{1i}(QPP Predictor_{ij}) + \epsilon_{ij}]
   \end{split}
\end{equation}
\looseness -1 Equation~\ref{eq_full_comp} moves from a 2-level specification towards a more compact version of all terms included in the resulting full model. 

Going further, we use and compare with two reduced models to calculate the Level 2 {\em variance components}, as outlined in Table 3. This helps to calculate the proportional gains in explained variance between nested models~\cite{maxwell2017designing,singer2003applied}. The reduced models can be obtained from $LME_{full}$ by setting one or more if its parameters to 0. We start with $LME_{average}$ that only assumes an average sARE value for the reference QPP predictor without a query type effect (written in composite form):
\begin{equation}\label{eq_umm}
sARE_{ij} = \gamma_{00} + \zeta_{0i} + \epsilon_{ij}
\end{equation}
Next, we obtain the second reduced model denoted as $LME_{QPP}$ in its composite form by adding the QPP measurement contribution as follows:
\begin{equation}\label{eq_ugm}
sARE_{ij} = \gamma_{00} + \gamma_{10}(QPP Predictor_{ij}) + \zeta_{0i} + \zeta_{1i}(QPP Predictor_{ij}) + \epsilon_{ij}
\end{equation}
The difference between $LME_{average}$ and $LME_{QPP}$ is that $LME_{QPP}$ assumes an effect of change per QPP predictor measurement. We obtain the proportional reduction in within-query variability as: $Pseudo-R^2_\epsilon = \frac{\sigma^2_{\epsilon_{LME_{average}}}-{\sigma^2_{\epsilon_{LME_{QPP}}}}}{\sigma^2_{\epsilon_{LME_{average}}}}$. The higher this proportion, the higher the reduction in within-query variability, and therefore, the more contribution the QPP predictor factor has in $LME_{QPP}$. Therefore, comparing $\sigma^2_\epsilon$ of $LME_{QPP}$ with that of $LME_{average}$ tells us how much of the total variability within queries can be attributed to QPP predictor. 

Similarly, the total variance explained by the Level 2 covariates (here: query type) can be found when comparing the Level-2 variances $\sigma^2_0$ and $\sigma^2_1$ of $LME_{full}$ with the ones of $LME_{QPP}$, since these two models only differ in the inclusion of the terms $\gamma_{01}(Query Type)$ and $\gamma_{11}(Query Type)$. 

\noindent Specifically, we have: $Pseudo-R^2_0 = \frac{\sigma^2_{0_{LME_{QPP}}}-{\sigma^3_{0_{LME_{full}}}}}{\sigma^2_{0_{LME_{QPP}}}}$ and $Pseudo-R^2_1 = \frac{\sigma^2_{1_{LME_{QPP}}}-{\sigma^3_{1_{LME_{full}}}}}{\sigma^2_{1_{LME_{QPP}}}}$. This tells us how much of the total variability between queries in initial status and rate of change, respectively, can be attributed to query type. The larger the reduction in the variance , the larger the contribution of query type at Level 2. The composite versions of $LME_{average}$ and $LME_{QPP}$ are, then, nested within Equation~\eqref{eq_full_comp} when setting one or more parameters to 0.

Based on the three defined LMEs, we address the following research questions:

\noindent {\bf RQ3} Is the accuracy of query performance influenced by query type more for dense retrieval than sparse retrieval?

\noindent {\bf RQ4} How important is the effect of differences in QPP predictors in predicting query performance?

\noindent {\bf RQ5} How important is the effect of differences in query type in predicting query performance?

We examine the resulting LMEs based on Equations~\ref{eq_full_comp},~\ref{eq_umm}, and ~\ref{eq_ugm}. We implement the proposed LMEs using the {\em lme4} R package~\cite{bates2009package,team2021r}, with Full Maximum Likelihood Estimation. Starting from $LME_{average}$, we sequentially add factors at Level 1 (and therefore, move to $LME_{QPP}$) and Level 2 ($LME_{full}$) if needed. At each step, we compare between the model that contains the added factor and the one that does not, and the decision is made based on the significance of fixed effects and the output value of Deviance~\cite{maxwell2017designing,singer2003applied} of each model, indicating the goodness-of-fit (the lower, the better). The deviance in this case is: $Deviance = -2LL_{Max}$, where $LL_{Max}$ is the maximised log-likelihood of each model. We report the final selected model in Table 4.

\subsection{Linear Mixed Model Results}
Table 4 shows the resulting LMEs for all retrieval methods on the sARE obtained by each of the three evaluation measures. In general, we notice that as the retrieval method becomes more effective, more terms are included in the LMEs, accounting for both the within and between-query variability. 

For TCT-ColBERT with $sARE_{MAP}$, we notice a significant within-query effect of QPP Predictor, and an interaction of predictor with query type (as indicated by the third term in the resulting LME's fixed effects). Specifically, we see two query types, Experience and Reason, as mentioned in Section~\ref{ssec:mod}. The interaction indicates that these query types perform better for some predictors, but worse for others. Indeed, Experience and Reason were found as {\em harder} questions in the original categorisation study~\cite{bolotova2022non}, so it is not surprising that for some predictors, they are worse. For $sARE_{NDCG}$, we only have an indication about a significantly large difference between QPP predictors, which is not affected by query type. For $sARE_{MRR}$, both QPP predictors and query types seem to behave similarly. 

For ANCE $sARE_{MAP}$, QPP predictors are sufficiently different (as seen from the inclusion of the second term in the corresponding LME), and are further influenced by Not a Question queries, also interacting with QPP predictors (inclusion of third and fourth fixed effects LME terms). For $sARE_{NDCG}$ and $sARE_{MRR}$, we don't observe any particular effect of query type, while the latter only shows a difference among predictors. Finally, BM25 sARE scores are overall less sensitive for between query differences, since query type is not included in any of the three evaluation measures resulting LMEs, (we only observe within-query differences in QPP predictor for $sARE_{MRR}$). In general, the two dense retrieval models are more sensitive to between-query differences than BM25; the effect of query type is exclusive to dense retrieval. To answer RQ3, the accuracy of query performance is influenced by query type more for dense retrieval than sparse retrieval.

\looseness -1 Next, we move on to the relative importance gains of both QPP predictors and query types, in order to address RQs 4 and 5. The top half of Table 5 shows the proportions of gained explained variance when adding Level 1 or Level 2 factors (with \xmark indicating no significant gains). For clarity, we include the bottom half of Table 5, which highlights the fixed effect terms included in each resulting LME (with \cmark indicating presence). To examine the increase in explained variance attributed to QPP predictor, we focus on the first row with $Pseudo-R^2_\epsilon$, which corresponds to models including the main effect of predictor. Specifically, for TCT-ColBERT and ANCE, a proportion of variance (from 12.4 to 14.6\%, and from 7.5\% to 16.5\%, respectively) is explained by differences in predictors. BM25 LMEs seem less affected by QPP predictor differences, since we only see a contribution for $sARE_MAP$. To answer RQ4, the effect of within-query differences due to QPP predictor are important for predicting query performance in the two dense retrieval models (more important compared to sparse retrieval), but still less important than the importance of query type (addressed below).

Finally, we examine the effect of query type on both intercepts and slopes. The row with $Pseudo-R^2_0$ corresponds to LMEs that include a main effect of query type. For ANCE, 17.2\% of the total variations between queries are due to differences in query type. This means that regardless of predictor, some types are more difficult to accurately predict effectiveness. In contrast, the variations due to main effect of query type are less prominent in TCT-ColBERT. Next, we move to the variances due to the interaction of query type with predictor, as indicated by the row with $Pseudo-R^2_1$. For both dense retrieval models, we see a very large proportion the variance in changes across predictors is attributed to query type. The variation due to the interaction effect is also present, but less noticeable for TCT-ColBERT on $sARE_{NDCG}$. No importance of query type is observed for BM25. To answer RQ5, query types are more important for $sARE_{MAP}$ compared to $sARE_{NDCG}$ and $sARE_{MRR}$ for dense retrieval models, while they are not influential for sparse retrieval.

\section{Conclusions}\pageenlarge{1}\label{sec:conc}
\looseness -1 We examined the accuracy of query performance predictors upon two single-representation dense retrieval methods (ANCE and TCT-ColBERT). In particular, we proposed new variants of existing unsupervised coherence-based predictors and managed to increase their performance in dense retrieval settings. In this way, we showed that changing the representation type from TF.IDF to neural embeddings provided by the dense retrieval models together with some further modifications is enough to generalise performance of unsupervised predictors in relation to supervised ones. Indeed, with increasing effectiveness brought by dense retrieval methods, our proposed predictors becomes more competitive, especially for NDCG@10 and MRR@10. At the same time, score-based predictors still remain very competitive and outperform supervised predictors, especially for MAP@100. Therefore, we prefer to allocate computational resources for ranking purposes, rather than predicting the effectiveness of a sparse ranker. 

Next, we examined differences among the identified predictors and find that they don't differ significantly in all cases. Importantly, we identify some unique characteristics of QPP in dense retrieval. Indeed, the type of query is responsible for a large amount of variance in query performance, and we can have query types that work well for some predictors but much worse for others. As a suggestion, it would be useful to examine the development of further predictors based on how they behave for more difficult query types.

\section*{Acknowledgements}
This work was supported by the UKRI Centre for Doctoral Training in Socially Intelligent Artificial Agents, Grant number EP/S02266X/1.

\bibliographystyle{ACM-Reference-Format}
\bibliography{sample-base}

\end{document}